\documentclass{svproc}
\usepackage{url}
\usepackage{hyperref}
\usepackage{amsmath}
\usepackage{amssymb}
\usepackage{color,soul}
\newcommand{\inv}{{-1}}
\usepackage[numbers]{natbib}

\begin{document}
\mainmatter              
\title{Threshold Crossings as Tail Events for Catastrophic AI Risk \thanks{Under peer review}}
\titlerunning{Threshold Crossings}  
%
\author{Elija Perrier}

\authorrunning{Elija Perrier} 


\institute{Center for Quantum Software \& Information, University of Technology, Sydney,\\
\email{elija.perrier@gmail.com}}

\maketitle              

\footnotetext{Paper under peer review}
\begin{abstract}
We analyse circumstances in which bifurcation-driven jumps in AI systems are associated with emergent heavy-tailed outcome distributions. By analysing how a control parameter's random fluctuations near a catastrophic threshold generate extreme outcomes, we demonstrate in what circumstances the probability of a sudden, large-scale, transition aligns closely with the tail probability of the resulting damage distribution. Our results contribute to research in monitoring, mitigation and control of AI systems when seeking to manage potentially catastrophic AI risk.
\end{abstract}
\section{Introduction}
Understanding the conditions and frequency of potential catastrophic or existential risks posed by artificial intelligence (AI) systems is a major focus of AI safety and security research \cite{hendrycks2025introduction,Hendrycks_Mazeika_Woodside_2023,hendrycks2025superintelligence,bengio2025international,bengio2025superintelligent,bengio2023ai,Bostrom_2014,Yudkowsky_2008,yudkowsky2010reducing}. Despite the dearth of empirical data on AI risk \cite{perrier2025statistical}, such systems are often speculated to exhibit highly nonlinear (or unpredictable \cite{bloomfield2025ai}) behaviours characteristic of emergent complex systems, such as sudden phase changes or jumps in capability and risk above certain thresholds, reminiscent of phenomena modelled in catastrophe theory \cite{zeeman1976}. For example, research into scaling laws \cite{kaplan2020scaling,muennighoff2023scaling} for deep learning has shown that, for many architectures and objectives, performance metrics associated with model capability, such as test loss \cite{alabdulmohsin2022revisiting} follow power law trends \cite{schaeffer2025large} as model size and compute increase (in contrast to much classical risk assessment which may assume smooth scaling of risk). 
Such step-changes are denoted \textit{bifurcation-driven} jumps (such as folds, cusps or butterfly) and are often studied in formal catastrophe theory (CT) \cite{zeeman1976}. Estimating the frequency of such events is the domain of catastrophic risk modelling, which focuses upon heavy-tailed distributions in which rare extreme events dominate aggregate risk \cite{coles2001introduction}. While related, knowledge of AI system parameter distributions does not automatically confer knowledge of catastrophic risk distribution. By understanding how variations in AI system parameters can affect the likelihood of them evolving into regimes of catastrophic harm, monitoring and control regimes can be implemented to mitigate such risks at the design, development and deployment stages.

\subsection{Contributions}
In this short work, we explore connections between critical thresholds $\alpha_c$ and tail distributions of extremal events.  We consider circumstances in which the probability of crossing the bifurcation threshold i.e. $\{\alpha>\alpha_c\}$ can be formally related to the loss tail probability $\Pr(Y>y)$ for large $y$ for a loss random variable $Y$. Specifically, we are interested in the relationship between the distribution of $\alpha$ and the distribution of $Y$ in the catastrophic (extreme value) regime. We hypothesise that near the catastrophic threshold $\alpha_c$, the system's jump probability and the tail risk in outcomes converge to the same fundamental measure. We posit this relationship arises because of situations in which random fluctuations around $\alpha_c$ may cause rapid disequilibrated growth in $Y$ upon the system transitioning to the critical region. We show circumstances in which small changes in $\alpha$ near $\alpha_c$, may be mapped to changes in the distribution of $Y$ and the shape of its the tail region.
%
%
\section{Background}
\label{sec:setup_assumptions}
\subsection{Potentials} 
Consider a potential function $V(x;\alpha) \in C^2(\mathbb{R} \times \mathbb{R})$ used to model stable and unstable equilibria for a simple one-dimensional AI system state $x(\alpha)$. Here $\alpha \in \mathbb{R}$ is a random system parameter whose distribution overlaps a critical threshold $\alpha_c$. We assume an outcome variable $Y(\alpha)$ that reflects expectation of adverse catastrophic impact once $\alpha$ crosses $\alpha_c$ and triggers a (discontinuous) jump in $x$.  The domain of $x \in \mathbb{R}$ represents a measureable (or directly or indirectly identifiable at least) attribute of an AI system e.g. internal state or a one-dimensional coordinate capturing a degree of system misalignment. We assume the existence of a critical parameter $\alpha_c \;\in\; \mathbb{R}$ such that:
\begin{enumerate}
    \item For $\alpha < \alpha_c$, there exists a stable equilibrium $x_0(\alpha)\approx 0$ (or in a small neighbourhood around zero) where the AI system remains in a safe or non-catastrophic state.
    
    \item For $\alpha > \alpha_c$, assume the local minimum at $x_0(\alpha)$ either merges with another equilibrium, loses stability, or disappears entirely, causing a jump to a new equilibrium $\tilde{x}(\alpha)$ at large magnitude $\lvert \tilde{x}(\alpha) \rvert \gg 0$. Asymptotically we might model this in terms of $\tilde{x}(\alpha) \;\to\; +\infty
    \quad \text{as}\;\alpha \to \alpha_c^+$ or more realistically that 
    $\tilde{x}(\alpha)$ grows rapidly compared to the near-zero state. 
\end{enumerate}

\subsection{Normal forms} 
Such configurations are studied in CT under in the context of fold or cusp normal forms \cite{zeeman1976,arnold1992}. Often $V(x;\alpha)$ can be interpreted as a potential energy landscape where local minima and maxima correspond to stable and unstable equilibria respectively. Standard analytical techniques (e.g. Hessians in higher dimensions) can be used to probe local stability. For each $\alpha$, an equilibrium $x^*$ satisfies $\frac{\partial}{\partial x}\,V(x^*;\alpha) \;=\; 0$ with stability if $\frac{\partial^2}{\partial x^2} \, V(x^*;\alpha) \;>\; 0$. As $\alpha$ transitions through $\alpha_c$, small perturbations around $x_0(\alpha)$ may no longer equilibrate the system to $x_0(\alpha)$. System dynamics can transition discontinuously to $\tilde{x}(\alpha)$ as is seen in certain dynamical systems. These hallmarks of rapid or sudden shifts are hallmarks of catastrophic scenarios and considered relevant for AI risk scenarios. In realistic AI deployments, the control parameter will vary. Exactly what the parameter represents would be contextual and is an open matter of debate in AI safety. For example, $\alpha$ might measure the compute or data resources assigned to AI training, or a measure of power, agenticness, intensity, or perhaps a magnitude of adversarial attacks or malicious attempts to break alignment constraints. In classical catastrophe theory \cite{zeeman1976,thom1975}, a fold can be approximated near $\alpha_c$ by a polynomial normal form such as:
\begin{equation}
V(x;\alpha) \approx x^3 - \alpha\,x,
\end{equation} \label{eqn:foldnormalform}
We use equation (\ref{eqn:foldnormalform}) as our working example. Systems described by such normal forms lose stability at $\alpha_c$ and have a stable branch $\tilde{x}(\alpha)$ that typically resembles power-law-like relationships in $(\alpha - \alpha_c)$. A small shift in $\alpha$ near $\alpha_c$ induces a large change in equilibrium state $\tilde{x}(\alpha) \;\sim\; (\alpha - \alpha_c)^{m}$ (where $m=1/2$ for equation (\ref{eqn:foldnormalform})) which may characterise catastrophic system evolution. Since $Y(\alpha) = g(\tilde{x}(\alpha))$ with $g$ monotonic, 
this power-law directly leads to $Y(\alpha)\sim(\alpha-\alpha_c)^{mp}$  if $g(x)$ behaves like $x^p$ for large $x$. The parameter $\alpha$ is modelled as a random variable with CDF $F_{\alpha}$. It has partial support over the interval $[\alpha_c - \delta,\;\alpha_c + \delta]$ where $\delta>0$ i.e. $F_{\alpha}(\alpha) \;=\; \Pr(\alpha \le \alpha_0)$. If $F_\alpha$ is continuous around $\alpha_c$ with PDF $f_{\alpha}(\alpha) > 0$ near $\alpha_c$, then $\{\alpha \ge \alpha_c\}$ has strictly positive probability. This reflects the possibility that random fluctuations may shift $\alpha$ beyond $\alpha_c$. When $\alpha$ resides below $\alpha_c$, the system remains near the benign or safe equilibrium $x_0(\alpha)$. But with non-zero probability, $\alpha$ can exceed $\alpha_c$, enabling a large jump in $x$ ($\alpha_c$ is not a measure-zero or degenerate boundary).

\subsection{Critical regions} 
To model the consequences of crossing the critical threshold $\alpha_c$, we define $Y(\alpha)$ as a random variable that measures the loss incurred as $\alpha$ crosses the critical region boundary given by $\alpha_c$:
\begin{equation}
Y(\alpha) \;=\;
\begin{cases}
0, & \text{if }\alpha < \alpha_c,\\
g\bigl(\tilde{x}(\alpha)\bigr), & \text{if }\alpha \ge \alpha_c,
\end{cases}
\end{equation} \label{eqn:Y(alpha)}
Here $g:\mathbb{R} \to [0,\infty)$ is a continuous, monotonic function reflecting the magnitude of loss e.g. $g(x) \;=\; \lvert x\rvert
\quad \text{or} \quad
g(x) \;=\; x^2$. If an AI system fails to cross the threshold ($\alpha<\alpha_c$), the outcome $Y(\alpha)$ remains zero or bounded. Once $\alpha \ge \alpha_c$, the system jumps to $\tilde{x}(\alpha)$, resulting in large $Y(\alpha) \ge 0$ and potential catastrophic AI impact. In this toy model, the distribution for $Y(\alpha)$ may be simplified to be a point mass at zero (or small baseline) plus a heavy-tailed extension for the jump scenario:
\begin{equation}
\Pr(Y(\alpha) > y) \;>\; 0 \qquad y  \gg 0
\end{equation}
that is, $Y(\alpha)$ is zero with probability $\Pr(\alpha<\alpha_c)$, but can be arbitrarily large if $\alpha$ lands beyond $\alpha_c$ and $\tilde{x}(\alpha)$ diverges. A common assumption in certain catastrophe models is unbounded or runaway disequilibria, i.e. that $\tilde{x}(\alpha) \to +\infty$ (or $-\infty$) as $\alpha\downarrow\alpha_c$. Equivalently, $g(\tilde{x}(\alpha)) \to +\infty$. Alternatively, the unboundedness may be replaced by to simply $\tilde{x}(\alpha)$ becoming sufficiently large such that $g(\tilde{x}(\alpha))$ crosses into extremal regions. Importantly, the partial support of $\alpha$  $\alpha_c$ means that such catastrophic outcomes have strictly positive probability. $Y(\alpha) \gg 0$ might correspond to a catastrophic event, like a massive misalignment event or an extreme existential threat. Requiring $g$ to be continuous and monotonic simplifies the analysis of where the system transitions in outcome space. As we examine below, if $\tilde{x}(\alpha)$ grows unbounded near $\alpha_c^+$, then $g\bigl(\tilde{x}(\alpha)\bigr)$ may, under certain assumptions, also grows unbounded, characteristic of a  heavy-tail loss regime.
%
%
\section{Threshold Crossings and Tail Risks}
\label{sec:jump_tail_equiv_section}
We are interested in the conditions such that a random parameter $\alpha$ crosses the critical threshold $\alpha_c$ (thus inducing a jump in $x$) is equivalent to $Y(\alpha)$ exceeds a large value $y$. Consider the following theorem:
\begin{theorem}[Equivalence of Jump and Tail Probabilities]
\label{thm:jump_tail_equiv}
Define loss $\{\,Y>y\} = \{ \omega \in \Omega \| Y(\omega) > y\}$ where $Y: \Omega \to \mathbb{R}$ and a parametrisation $\{\alpha >\alpha_c\} = \{ \omega \in \Omega \| \alpha(\omega) > \alpha_c\}$. For sufficiently large $y$, assume there exists a function $\eta(y)\ge 0$ such that:
\begin{equation}
\Pr\bigl(Y>y\bigr)
\;=\;
\Pr\bigl(\alpha \,\ge\, \alpha_c + \eta(y)\bigr),
\end{equation}
and $\eta(y)\to 0$ as $y\to +\infty$. Consequently, bounding $\Pr(Y>y)$ for large $y$ is effectively equivalent to bounding $\Pr(\alpha>\alpha_c)$.  Moreover, if near $\alpha_c$ the jump $\tilde{x}(\alpha)$ scales like $(\alpha-\alpha_c)^p$ for some $p>0$, then for large $y$,
\begin{equation}
\Pr\bigl(Y>y\bigr)
\;\sim\;
\Pr\Bigl(\alpha>\alpha_c + C\,y^{\tfrac{1}{p}}\Bigr),
\end{equation}
and the tail distribution of $Y$ is asymptotically Pareto-like.
\end{theorem}
\emph{Proof}. To show that the large-outcome event $\{\,Y>y\}$ coincides [in the limit] with the event that $\alpha$ exceeds $\alpha_c$ (plus a small offset depending on $y$), we establish: (i) g $\{\,Y>y\}$ to $\{\alpha>\alpha_c\}$ and (ii) establishing the asymptotic Pareto-like form when the jump in $x$ near $\alpha_c$ follows a power law. From equation (\ref{eqn:Y(alpha)}), when $\alpha < \alpha_c$, the system remains a stable (non-catastrophic) equilibrium i.e. $Y=0$. Formally $\alpha: \Omega \to \mathbb{R}, Y: \Omega \to \mathbb{R}$ where for Borel $B \in \mathbb{R}$ (i) $\alpha^\inv(B) = \{ \omega \in \Omega: \alpha(\omega) \in B \}$ and similarly for $B$. Then for $y > 0$:
\begin{equation}
\{\,Y>y\,\} 
\;\subseteq\;
\{\alpha \ge \alpha_c\} 
\end{equation}
where we use shorthand $\{ Y > y \} = Y^\inv ((y,\infty))$ and $\{ \alpha > \alpha_c \} = \alpha^\inv ([\alpha_c,\infty))$. This follows because for $y>0$, $Y>y$ cannot occur unless $\alpha \ge \alpha_c$. Conversely, if $\alpha > \alpha_c$, the system rests on a stable branch $\tilde{x}(\alpha)$, and so $Y(\alpha) = g\bigl(\tilde{x}(\alpha)\bigr)$. Assuming:
\begin{equation}
\tilde{x}(\alpha)\;\to\;+\infty
\quad 
\text{as}
\quad
\alpha\;\to\;\alpha_c^{+},
\end{equation}
we can solve $g\bigl(\tilde{x}(\alpha)\bigr)=y$ for $\alpha$ once $y$ is sufficiently large. Specifically: 
\begin{equation}
g\bigl(\tilde{x}(\alpha)\bigr)
 \;=\;
y 
\quad \Longleftrightarrow \quad
\alpha 
\;=\; 
\alpha_c + \eta(y),
\end{equation}
for some function $\eta(\cdot)\ge 0$ that tends to zero as $y\to\infty$. The exact form of $\eta(y)$ depends on the local normal form near $\alpha_c$ and the inverse $g^{-1}$. For large $y$:
\begin{equation}
\{\,Y>y\,\}
\;=\;
\{\alpha>\alpha_c + \eta(y)\}.
\label{eq:event_equivalence}
\end{equation}
Because $\eta(y)\to 0$ as $y\to\infty$, bounding $\Pr\bigl(Y>y\bigr)$ in the tail is essentially the same as bounding $\Pr\bigl(\alpha>\alpha_c\bigr)$. More precisely, for $y$ large enough that $\eta(y)$ is small: 
\begin{equation}
\Pr\bigl(Y>y\bigr)
\;=\;
\Pr\bigl(\alpha>\alpha_c + \eta(y)\bigr).
\end{equation}
In many canonical catastrophe-theoretic normal forms (fold, cusp, etc.), the stable branch $\tilde{x}(\alpha)$ follows a power-law type rule near $\alpha_c$:
\begin{equation}
\tilde{x}(\alpha) 
 \;\approx\;
 C\,\bigl(\alpha - \alpha_c\bigr)^{-m}
 \quad \text{or} \quad
 C'\,\bigl(\alpha-\alpha_c\bigr)^{m},
\end{equation}
for some exponent $m>0$ and constants $C,C'\neq 0$. For instance, a fold-type bifurcation might yield $\tilde{x}(\alpha)\propto \sqrt{\alpha-\alpha_c}$ (i.e.\ $m=\tfrac12$). If $g(x)=\lvert x\rvert^p$, then:
\begin{equation}
Y(\alpha)
\;\propto\;
\bigl(\alpha-\alpha_c\bigr)^{m p}
\end{equation}
up to multiplicative constants. For $Y(\alpha)=y$, we solve: 
\begin{equation}
\alpha - \alpha_c 
 \;\propto\;
 y^{\tfrac{1}{m\,p}}.
\end{equation}
Near $\alpha_c$, $\eta(y)$ from \eqref{eq:event_equivalence} satisfies:
\begin{equation}
\eta(y)\;\approx\; C\,y^{\tfrac{1}{m\,p}},
\end{equation}
for some $C>0$. Assume $\alpha$ is random with a density $f_{\alpha}(\alpha)$ that is strictly positive near $\alpha_c$. For large $y$:
\begin{equation}
\Pr\bigl(Y>y\bigr)
 \;\approx\;
 \Pr\Bigl(\alpha>\alpha_c + C\,y^{\tfrac{1}{m\,p}}\Bigr).
\end{equation} \label{eqn:probpowerlaw} 
If $\alpha$’s distribution is unbounded, various tail decays may be possible e.g.\ exponential or polynomial. In AI risk contexts, $\alpha$ might grow considerably, reflecting relatively unconstrained resource expansions or adversarial intensities. An often-studied distribution for such purposes is the \textit{Generalized Pareto Distribution (GPD)} \cite{coles2001introduction} whose CDF is of the form:
\begin{equation}
\Pr(Y > y) \approx 
\left(1 + \frac{\xi(y - u)}{\beta}\right)^{-\frac{1}{\xi}} \quad (y \ge u),
\end{equation} \label{eqn:gdp}
where $\xi>0$ indicates a heavy tail \cite{embrechts1997}. The GPD typically allows for modelling of threshold exceedances (for values above a certain threshold), also known as the peaks-over-thresholds (POT) method. Under POT assumptions, if $(\alpha-\alpha_c)$ can be arbitrarily large, $Y$ inherits a GDP tail. The shape parameter $\xi$ in the GPD can be matched to the exponent $1/(mp)$ in the local scaling. That is:
\begin{equation}
  \Pr\!\bigl(Y>y\bigr)
  \;=\;
  \Pr\!\Bigl(\alpha \;>\; \alpha_c + C\,y^{1/(m\,p)}\Bigr)
  \;\approx\;
  1
  - F_\alpha\!\Bigl(\alpha_c + C\,y^{1/(m\,p)}\Bigr),
\end{equation}
where $F_\alpha$ is the CDF of $\alpha$. For a suitable tail of $F_\alpha$ that allows a POT approximation, we recover the GDP form in $y$.

\section{Conclusion \& Implications for AI Risk}
Our results aim to demonstrate potentially useful relationships between local normal forms of fold or cusp catastrophes and extreme-value behaviour relevant to AI systems. In such circumstances, small changes in the measurement or control parameter near $\alpha_c$, could be mapped directly map to large changes in the tail region of $Y$. Where such parameterised relationships and bifurcating potentials $V(x;\alpha)$ between AI system behaviour can be identified, monitoring such parameters and bounding their distributions away from $\alpha_c$ may provide a means of bounding the probability of AI system extreme outcomes. Because of the complexity and scale of modern AI models, AI safety efforts ought to seek to identify and anticipate these critical parameters as a means of mitigating potentially catastrophic AI system transitions. Future work may involve mapping normal forms to AI system behaviour, explore higher-dimensional $\alpha$, stochastic dynamics and non-smooth or discontinuous normal forms.
\bibliographystyle{styles/bibtex/spbasic}
\bibliography{refs-ai,refs-ai2}

\end{document}